# Bursting Money Bins
## The ice and water structure

Franco Bagnoli,
Dept. of Physics and Astronomy and Center for the Study of Complex Dynamics, University of Florence, Italy
Via G. Sansone, 1 50019 Sesto Fiorentino (FI) Italy
franco.bagnoli@unifi.it

In the classic comics by Carl Barks, "The Big Bin on Killmotor Hill" [1], Uncle Scrooge, trying to defend his money bin from the Beagle Boys, follows a suggestion by Donald Duck, and fills the bin with water. Unfortunately, that night is going be the coldest one in the history of Ducksburg. The water freezes, bursting the ``ten-foot walls'' of the money bin, and finally the gigantic cube of ice and dollars slips down the hill up to the Beagle Boys lot.

That water expands when freezing is a well-known fact, and it is at the basis of an experiment that is often involuntary performed with beer bottles in freezers. But why does the water behave this way? And, more difficult, how can one illustrate this phenomenon in simple terms?

First of all, we have to remember that the temperature is related to the kinetic energy of molecules, which tend to stay in the configuration of minimal energy. In general, if one adopts a simple ball model for atoms, the configuration of minimal energy is more compact than a more energetic (and thus disordered) configuration. But this is not the case for water.

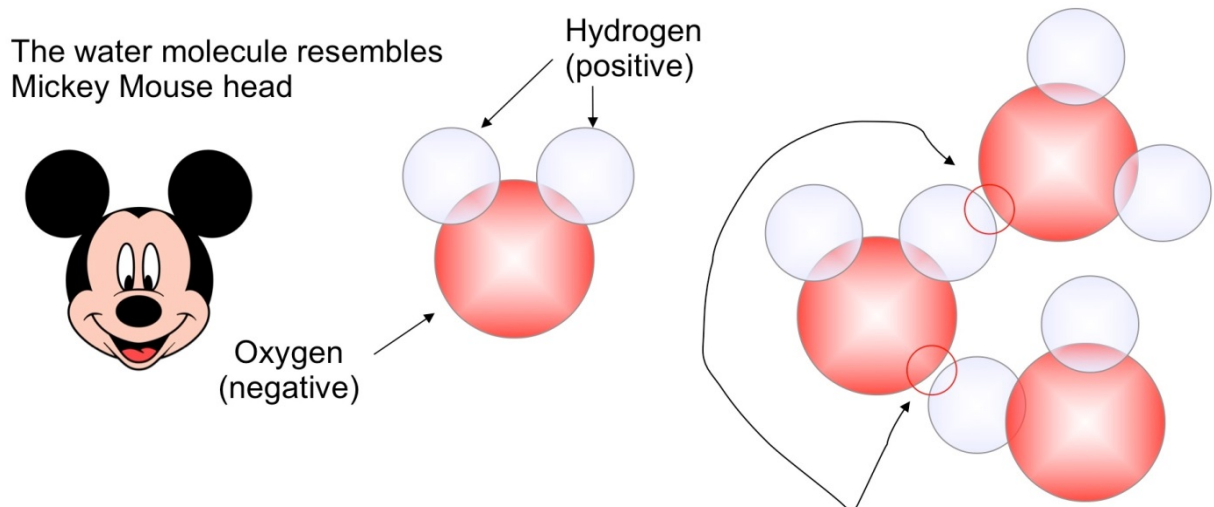

FIG. 1: A Schematic representation of the water molecule

Indeed, water molecules resemble the head of Mickey Mouse (see Fig. 1), the two hydrogen atoms being the ears. It is difficult to draw the crystal structure of water, since it is intrinsically three-dimensional, but one can make use of the ``Mercedes Benz'' model [2].

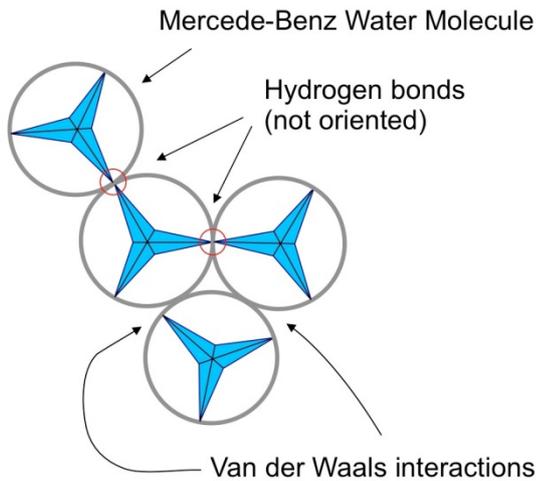

FIG. 2: Water molecules in the Mercedes-Benz model

In this model, the water molecules are represented as the famous Mercedes-Benz symbol (see Fig. 2), and the minimum energy is obtained when two spikes join (an unoriented version of the hydrogen bond). It is not difficult to make some cardboard Mercedes-Benz molecules and let people play with them.

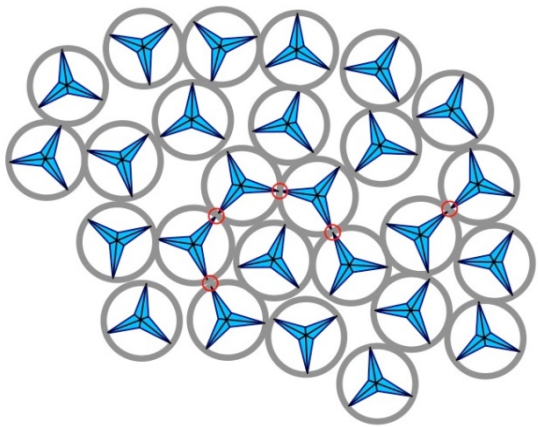
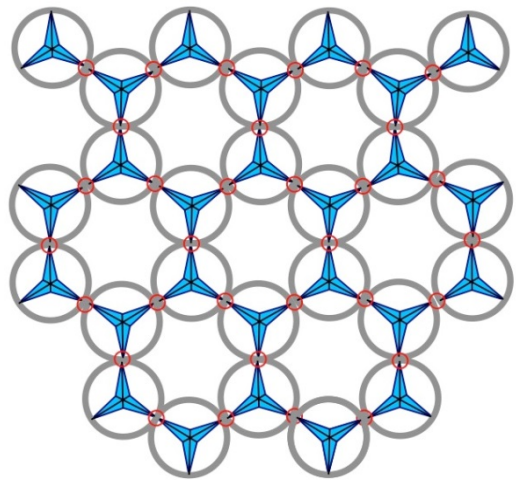

FIG. 3: Water and ice in the Mercedes-Benz model

By means of this model it is now easy to illustrate that the minimum energy configuration contains a lot of ``void'' (see right-hand side of Fig. 3), and this explains the lower density of ice with respect to water. It may also serve to illustrate the phenomenon of regelation (melting under pressure and freezing again when the pressure is reduced).

However, in order to have a stable ice structure, we first have to reduce the kinetic energy of water molecules, and then remove also the difference in energy from the Van der Waals to the hydrogen bonds, i.e., we have to extract heat. In macroscopic terms, the first is called the heat capacity of water (specific heat: $C$ = 4.187 kJ/kg K), and the second the latent heat of melting ($l$

= 334 kJ/kg). If one wants to further lower the temperature of ice, one has also to extract the energy corresponding to the specific heat of ice (2.108 kJ/kg K).

In Carl Barks' comics, Uncle Scrooge often states that his money bin is "three cubic acres big", which is obviously nonsense, because the acre is a surface unit. If we assume that each face of the cube is an acre (around 4,000 m$^2$), we have a side of about 60 m, and a volume $V$ of about 200,000 m$^3$. I was unable to find the density of a random packing of coins (it seems that only spheres and M&M's have been studied [3]), so I arranged a small experiment with 174 ten cents coins. The resulting packing density is about 0.55, much less than spheres (0.64), but there are pronounced finite-size effects[1]. The total amount of water is thus $W$ = 0.45×2×10$^5$×10$^3$ = 9×10$^7$ kg. Assuming that initially the money bin (and the water inside) is at $T$ = 10°C, the total amount of heat to be removed to obtain ice at 0°C is $W \times (C \times T + I)$ = 3.4×10$^{13}$ J if we disregard the heat capacity of coins – that of copper/brass is a mere 0.4 kJ/kg K – and that of the walls, about 0.8 kJ/kg K. If the freezing occurs in, say, 5 hours (disregarding the time required for the heat conduction), this means a cooling capacity of 113 GW! Indeed, to freeze water is not an easy task, and it is a common practice to put cans full of water near fruit trees in order to protect them from frost.

Finally, we close with a challenge: what is the temperature at the bottom of the sea?[2]

---

[1] This allows us to compute how rich Scrooge McDuck is. The volume of a 1 dollar coin is about 1.1 cm$^3$, and considering a 0.55 packing density it occupies an average volume of 2 cm$^3$. Dividing $V$ by this number we obtain about 10$^{11}$ dollars, one hundred thousand millions or one hundred billions.

[2] Let us first analyse the case of fresh water (a deep lake). The maximum density of water, according to our model, is somewhat above zero (in real water it is about 4 °C). Below this temperature, ice-like structures (ruled by hydrogen bonds) are common, lowering the density towards that of ice. Above this temperature the Van Waals interactions dominate and the water behaves more like a ``standard'' liquid. In the presence of salt things change a bit, the freezing point lowers and so does the maximum water density. So the bottom of the ocean is about 3 °C.